# Design of AlGaAs Laser Power Converters for the First Transmission Window

Marina Delgado, Iván García, Carlos Algora
Instituto de Energía Solar, Universidad Politécnica de Madrid
ETSI Telecomunicación, Avenida Complutense, 30 – Madrid (SPAIN)-
marina.delgado@ies.upm.es

***Abstract:*** *Thermalization losses in Laser Power Converters for the first transmission window occur due to the energy difference between 808 nm wavelength photons and the GaAs bandgap energy which results in lower efficiency. In this work, we analyze the potential of increasing the bandgap using AlGaAs to minimize the energy difference. The optimum aluminum content is 7 % in both single and triple junction converters reaching a maximum ideal efficiency of 71% at 1 W/cm$^2$ and 77% at 50 W/cm$^2$.*

## 1. Introduction

Power-by-light systems are used to power equipment in exclusion zones where the electricity is not adequate due to the risk involved. These systems consist of a monochromatic light source, an optical fiber and a photovoltaic power converter. The light, usually coming from a laser, travels through the fiber until it reaches the converter where it is transformed into electricity [1].

The typical emission wavelength used for the first transmission window is 808 nm due to the inexpensive high-power light sources availability. Nevertheless, it does not allow maximum efficiency in GaAs power converters due to the thermalization losses associated to the difference between the photons and the GaAs bandgap energy.

For this reason, in this work we present a theoretical analysis which consists of increasing the absorbent bandgap, using AlGaAs in single (SJ) and triple junction (3J) converters looking for the optimum aluminum composition (Al%).

## 2. Theoretical model

We obtain I-V curves using the single-diode model from different ad-hoc software tools programmed in the Igor Pro environment. These tools are purely 1D so the distributed series resistance effect is not considered.

To know the short circuit current ($J_{sc}$) we use a tool that simulates the external quantum efficiency based on the "Generalized Matrix Method" [2]. We assume an absorber with collection efficiency equal to 1, which is a good approximation in GaAs cells as shown by the experimental data previously obtained at IES-UPM. We also assume that laser radiation is spectrally narrow enough to approach an emission line at a wavelength of 808 nm, considering 1 W/cm$^2$ and 50 W/cm$^2$ as input power densities.

Steiner et al. model is used to obtain the open-circuit voltage ($V_{oc}$). Because of GaAs cells can be considered homogeneous with negligible recombination at the interfaces, a single parameter aggregating the effects of recombination, namely the internal luminescence efficiency ($\eta_{int}$) is used, which get values of 1 and 0.5 to consider ideal and more real efficiencies, respectively. With $\eta_{int}$ and the optical parameters of the structure we obtain the amount of photons that can escape through the frontal surface, the external luminescence efficiency ($\eta_{ext}$), and the total $V_{oc}$ [3] considering also the ideal $V_{oc}$ calculated in the detailed balance limit ($V_{db}$).

Regarding the power converter structure, the multi junction (MJ) photovoltaic cell is shown in Fig. 1 [4, 5]. The SJ cells simulated have the equivalent structure without tunnel junctions. MgF$_2$ and ZnS antireflective (AR) layers are used. Both AR and absorbent layers thicknesses for each Al% are optimized. SJ absorber was adjusted so that the photocurrent was always 99.7% of the maximum achievable setting a limit in 4000 nm. The bottom subcell of the MJ is kept at 3500 nm to facilitate the calculations.

| AR layer | MgF$_2$ | X |
|---|---|---|
| AR layer | ZnS | X |
| Adhesion AR layer | MgF$_2$ | 3 nm |
| Window SB1 | Ga$_{0.5}$In$_{0.5}$P | 300 nm |
| Absorber SB1 | Al$_x$Ga$_{1-x}$As | X |
| BSF SB1 | Al$_{0.4}$Ga$_{0.6}$As | 200 nm |
| Anode TJ1 | Al$_{0.4}$Ga$_{0.6}$As | 50 nm |
| Cathode TJ1 | Ga$_{0.5}$In$_{0.5}$P | 25 nm |
| Window SB2 | Ga$_{0.5}$In$_{0.5}$P | 25 nm |
| Absorber SB2 | Al$_x$Ga$_{1-x}$As | X |
| BSF SB2 | Al$_{0.4}$Ga$_{0.6}$As | 200 nm |
| Anode TJ2 | Al$_{0.4}$Ga$_{0.6}$As | 50 nm |
| Cathode TJ2 | Ga$_{0.5}$In$_{0.5}$P | 25 nm |
| Window SB23 | Ga$_{0.5}$In$_{0.5}$P | 25 nm |
| Absorber SB3 | Al$_x$Ga$_{1-x}$As | 3500 nm |
| BSF SB3 | Al$_{0.4}$Ga$_{0.6}$As | 200 nm |
| Substrate | GaAs | - |

Fig. 1. Triple junction power converter structure.

Finally, the optical parameters of the materials are obtained from the literature and our own experimental data. For AlGaAs at different compositions we use the Adachi model [6]. We assume that the minority carrier parameters do not change significantly within the small Al% we swept [7].

## 3. Results

The optimum Al% in AlGaAs is 7% for single and triple junction, with a 77 % ideal efficiency ($\eta_{int}$=1) in both cases at 50 W/cm$^2$. However, for a realistic case with lower $\eta_{int}$, an efficiency loss is observed, which is higher in SJ converters as shown in Fig.2.

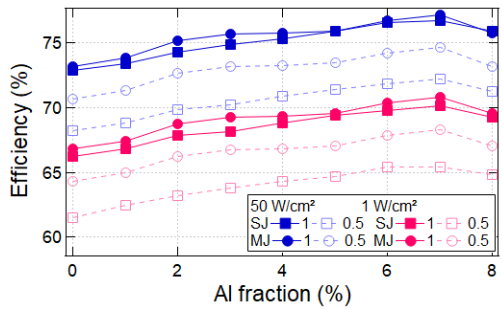

Fig. 2. The efficiency as a function of Al% for single and triple junction, $P_{in}$=1 W/cm$^2$ and 50 W/cm$^2$, $\eta_{int}$ =1 and $\eta_{int}$ =0.5.

The parameter that causes the reduction in efficiency as the material quality gets worse is the $V_{oc}$ through a lower $\eta_{ext}$. To understand why the SJ efficiency is more affected by the lower quality, we first show a contour plot of the $\eta_{ext}$ as the thickness and $\eta_{int}$ change (Fig.3). For thinner thicknesses, the impact of $\eta_{int}$ on $\eta_{ext}$ is significantly reduced. Since in a MJ the top subcells have reduced absorber thicknesses (around 500 nm), the total $V_{oc}$ is more immune to variations in $n_{int}$. Therefore, SJ and the bottom cell of a multi junction are the ones that suffer the most this effect.

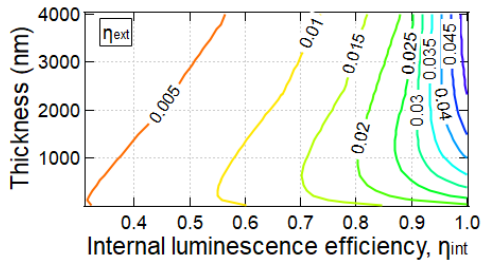

Fig. 3. $\eta_{ext}$ as a function of Al$_{0.07}$Ga$_{0.93}$As thickness and the $\eta_{int}$ at 1 W/cm$^2$ in a single junction power converter.

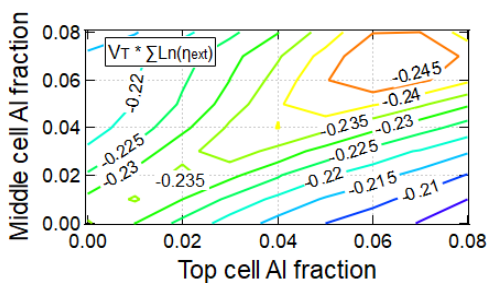

Fig. 4. $V_{oc}$ term: $\Sigma V_T*Ln(\eta_{ext})$, of the triple junction as a function of the Al% in Top/Middle subcell (Al$_{0.07}$Ga$_{0.93}$As in bottom cell) for $\eta_{int}$=1 and $P_{in}$ =1 W/cm$^2$.

On the other hand, modification of the optical environment for each subcell as the Al% changes creates interesting effects on the $V_{oc}$. We observe that a higher Al% in the top cell decreases its $\eta_{ext}$ due to a higher relative absorption in the upper layers in the structure, but it also slightly increases the $\eta_{ext}$ of the middle cell, and vice versa. The overall effect on the $V_{oc}$ is plotted in Fig. 4, showing a minimum of the $V_{oc}$ "loss" term due to $\eta_{ext}$ at about the optimal Al% in terms of $V_{oc}$ and efficiency. The $V_{oc}$ evolution is dominated by the change in bandgap (through the $V_{db}$ term in the model used). However, when implementing these AlGaAs-based PC, the variations of the internal optics of the whole semiconductor structure must be taken into account to prevent assigning unexpected $V_{oc}$ variations to effects other than variations in the optical environment.

## 4. Conclusions

We consider using AlGaAs as absorber to reduce the loss of efficiency caused by an energy difference between 808 nm laser photons and the GaAs bandgap of the converter. We find that 7 % is the optimum Al% in AlGaAs for single and triple junction, with a 77 % ideal efficiency at 50 W/cm$^2$ input power. For realistic cases with $\eta_{int}$ < 1 we observe a larger drop in efficiency for the single junction cells, which can be explained by a stronger decrease in $V_{oc}$ in its thicker absorber as compared to the thin upper cells in the triple junction devices.

We have studied the complex effect of variations in the internal optics as the Al% changes in the subcells. Although the $V_{oc}$ trend is dominated by the effect of the bandgap on the detail-balance voltage, understanding the variations in the internal optics is important to accurately guiding the experimental implementation of these power converters.

**Acknowledgments**

This project has been funded by the Comunidad de Madrid with the project with reference Y2018/EMT-4892 (TEFLON-CM) and by Universidad Politécnica de Madrid by Programa Propio. I. García is funded by the Spanish Programa Estatal de Promoción del Talento y su Empleabilidad through a Ramon y Cajal grant (RYC-2014- 15621).